\documentstyle[twoside]{article}

\catcode`\@=11
\long\def\@makefntext#1{ 
\protect\noindent \hbox to 3.2pt {\hskip-.9pt
$^{{\eightrm\@thefnmark}}$\hfil}#1\hfill} 

\def\thefootnote{\fnsymbol{footnote}}
 \def\@makefnmark{\hbox to 0pt{$^{\@thefnmark}$\hss}}  

\def\ps@myheadings{\let\@mkboth\@gobbletwo
\def\@oddhead{\hbox{} 
\rightmark\hfil\eightrm\thepage}
\def\@oddfoot{}\def\@evenhead{\eightrm\thepage\hfil 
\leftmark\hbox{}}\def\@evenfoot{}
\def\sectionmark##1{}\def\subsectionmark##1{}}

\oddsidemargin=\evensidemargin
\renewcommand{\thefootnote}{\fnsymbol{footnote}}

\newcounter{sectionc}\newcounter{subsectionc}\newcounter{subsubsectionc}
\renewcommand{\section}[1] {\vspace{12pt}\addtocounter{sectionc}{1}
\setcounter{subsectionc}{0}\setcounter{subsubsectionc}{0}\noindent
        {\tenbf\thesectionc. #1}\par\vspace{5pt}}
\renewcommand{\subsection}[1] {\vspace{12pt}\addtocounter{subsectionc}{1}
        \setcounter{subsubsectionc}{0}\noindent
        {\bf\thesectionc.\thesubsectionc. {\kern1pt \bfit #1}}\par\vspace{5pt}}
\renewcommand{\subsubsection}[1]
{\vspace{12pt}\addtocounter{subsubsectionc}{1}
        \noindent{\tenrm\thesectionc.\thesubsectionc.\thesubsubsectionc.
        {\kern1pt \tenit #1}}\par\vspace{5pt}}
\newcommand{\nonumsection}[1] {\vspace{12pt}\noindent{\tenbf #1}
        \par\vspace{5pt}}

\newcounter{appendixc}
\newcounter{subappendixc}[appendixc]
\newcounter{subsubappendixc}[subappendixc]
\renewcommand{\thesubappendixc}{\Alph{appendixc}.\arabic{subappendixc}}
\renewcommand{\thesubsubappendixc}
        {\Alph{appendixc}.\arabic{subappendixc}.\arabic{subsubappendixc}}

\renewcommand{\appendix}[1] {\vspace{12pt}
        \refstepcounter{appendixc}
        \setcounter{figure}{0}
        \setcounter{table}{0}
        \setcounter{lemma}{0}
        \setcounter{theorem}{0}
        \setcounter{corollary}{0}
        \setcounter{definition}{0}
        \setcounter{equation}{0}
        \renewcommand{\thefigure}{\Alph{appendixc}.\arabic{figure}}
        \renewcommand{\thetable}{\Alph{appendixc}.\arabic{table}}
        \renewcommand{\theappendixc}{\Alph{appendixc}}
        \renewcommand{\thelemma}{\Alph{appendixc}.\arabic{lemma}}
        \renewcommand{\thetheorem}{\Alph{appendixc}.\arabic{theorem}}
        \renewcommand{\thedefinition}{\Alph{appendixc}.\arabic{definition}}
        \renewcommand{\thecorollary}{\Alph{appendixc}.\arabic{corollary}}
        \renewcommand{\theequation}{\Alph{appendixc}.\arabic{equation}}
        \noindent{\tenbf Appendix \theappendixc #1}\par\vspace{5pt}}
\newcommand{\subappendix}[1] {\vspace{12pt}
        \refstepcounter{subappendixc}
        \noindent{\bf Appendix \thesubappendixc. {\kern1pt \bfit #1}}
        \par\vspace{5pt}}
\newcommand{\subsubappendix}[1] {\vspace{12pt}
        \refstepcounter{subsubappendixc}
        \noindent{\rm Appendix \thesubsubappendixc. {\kern1pt \tenit #1}}
        \par\vspace{5pt}}

\newcommand{\textlineskip}{\baselineskip=13pt}
\newcommand{\smalllineskip}{\baselineskip=10pt}

\def\eightcirc{
\begin{picture}(0,0)
\put(4.4,1.8){\circle{6.5}}
\end{picture}}
\def\eightcopyright{\eightcirc\kern2.7pt\hbox{\eightrm c}}

\newcommand{\copyrightheading}[1]
        {\vspace*{-2.5cm}\smalllineskip{\flushleft
        {\eightrm Modern Physics Letters A, #1}\\
        {\eightrm $\eightcopyright$\, World Scientific Publishing
         Company}\\
         }}

\newcommand{\publisher}[2]{{\begin{center}\eightrm\smalllineskip
        Received #1\\
        Revised #2
        \end{center}
        }}

\def\abstracts#1#2#3{{
        \centering{\begin{minipage}{4.5in}\baselineskip=10pt\eightrm
        \centerline{ABSTRACT}
        \parindent=0pt #1\par
        \parindent=15pt #2\par
        \parindent=15pt #3
        \end{minipage} }\par}}

\renewenvironment{thebibliography}[1]                   
        {\ninerm
         \baselineskip=11pt                             
         \begin{list}{\arabic{enumi}.}
        {\usecounter{enumi}\setlength{\parsep}{0pt}
         \setlength{\leftmargin 17pt}{\rightmargin 0pt} 
         \setlength{\itemsep}{0pt} \settowidth          
        {\labelwidth}{#1.}\sloppy}}{\end{list}}

\newcounter{itemlistc}
\newcounter{romanlistc}
\newcounter{alphlistc}
\newcounter{arabiclistc}

\newcommand{\fcaption}[1]{
        \refstepcounter{figure}
        \setbox\@tempboxa = \hbox{\eightrm Fig.~\thefigure. #1}
        \ifdim \wd\@tempboxa > 5in
           {\begin{center}
        \parbox{5in}{\eightrm \smalllineskip Fig.~\thefigure. #1 }
            \end{center}}
        \else
             {\begin{center}
             {\eightrm Fig.~\thefigure. #1}
              \end{center}}
        \fi}

\newcommand{\tcaption}[1]{
        \refstepcounter{table}
        \setbox\@tempboxa = \hbox{\eightrm Table~\thetable. #1}
        \ifdim \wd\@tempboxa > 5in
           {\begin{center}
        \parbox{5in}{\eightrm\smalllineskip Table~\thetable. #1 }
            \end{center}}
        \else
             {\begin{center}
             {\eightrm Table~\thetable. #1}
              \end{center}}
        \fi}

\def\@citex[#1]#2{\if@filesw\immediate\write\@auxout    
        {\string\citation{#2}}\fi                       
\def\@citea{}\@cite{\@for\@citeb:=#2\do                 
        {\@citea\def\@citea{,}\@ifundefined             
        {b@\@citeb}{{\bf ?}\@warning
        {Citation `\@citeb' on page \thepage \space undefined}}
        {\csname b@\@citeb\endcsname}}}{#1}}

\newif\if@cghi
\def\cite{\@cghitrue\@ifnextchar [{\@tempswatrue
        \@citex}{\@tempswafalse\@citex[]}}
\def\citelow{\@cghifalse\@ifnextchar [{\@tempswatrue
        \@citex}{\@tempswafalse\@citex[]}}
\def\@cite#1#2{{$\null^{#1}$\if@tempswa\typeout
        {IJCGA warning: optional citation argument
        ignored: `#2'} \fi}}

\def\pmb#1{\setbox0=\hbox{#1}
        \kern-.025em\copy0\kern-\wd0
        \kern.05em\copy0\kern-\wd0
        \kern-.025em\raise.0433em\box0}

\def\fnt#1#2{\footnotetext{\kern-.3em
        {$^{\mbox{\scriptsize #1}}$}{#2}}}

\def\fpage#1{\begingroup
\voffset=.3in
\thispagestyle{empty}\begin{table}[b]\centerline{\footnotesize #1}
        \end{table}\endgroup}

\def\runninghead#1#2{\pagestyle{myheadings}
\markboth{{\eightit{\quad #1}}\hfill}{\hfill{\eightit{#2\quad}}}}

\font\tenbf=cmbx10
\font\tenit=cmti10
\font\tenit=cmti10
\font\bfit=cmbxti10 at 10pt
 1
 1
 1

\font\ninerm=cmr9

\font\eightrm=cmr8
\font\eightit=cmti8

\newtheorem{theorem}{\indent Theorem}

\newtheorem{lemma}{Lemma}

\runninghead{Operator algebra
$\ldots$} {Operator algebra
$\ldots$}

\def\ffM{ \hbox{$M$}\kern-.9em\hbox{$\overline{\phantom{N}}$}}
\def\p{\partial}
\def\half{{\textstyle{1\over2}}} 
\def\al{\alpha}
\def\be{\beta}

\def\de{\delta}  \def\De{\Delta}

\def\th{\theta}

\def\la{\lambda} \def\La{\Lambda}
\def\rh{\rho}

\def\ph{\phi}

\def\cA{{\cal A}}
\def\cC{{\cal C}}
\def\cF{{\cal F}}
\def\cH{{\cal H}}
\def\cI{{\cal I}}
\def\cL{{\cal L}}
\def\cO{{\cal O}}
\def\cP{{\cal P}}
\def\cR{{\cal R}}
\def\cS{{\cal S}}
\def\cV{{\cal V}}
\def\cW{{\cal W}}
\def\mapright#1{\smash{\mathop{\longrightarrow}\limits^{#1}}}
\def\ZZ{{Z\!\!\!Z}}
\def\CC{{I\!\!\!\!C}}
\def\bfg{{\bf g}}
\def\bfh{{\bf h}}

\begin{document}
\normalsize\textlineskip
{\thispagestyle{empty}
\setcounter{page}{1}

\renewcommand{\thefootnote}{\fnsymbol{footnote}} 

\copyrightheading{Vol. 0, No. 0 (1992) 000--000}

\vspace*{0.88truein}

\fpage{1}
\centerline{\bf CHIRAL RING IN THE $4D$ $\cW_3$ STRING}
\vspace{0.4truein}
\centerline{\footnotesize J. McCarthy}
\vspace*{0.015truein}
\centerline{\footnotesize\it Department of Physics and
Mathematical Physics, University of
Adelaide}
\baselineskip=10pt
\centerline{\footnotesize\it Adelaide SA 5005, AUSTRALIA.}
\vspace{0.225truein}
\publisher{(received date)}{(revised date)}

\vspace*{0.21truein}
\abstracts{\noindent
 }{I summarize some recent results obtained in collaboration with
P.~Bouwknegt and K.~Pilch on the spectrum of physical states
in $\cW_3$ gravity coupled to $c=2$ matter. In particular, it is
shown that the algebra of operators corresponding
to physical states -- defined as a semi-infinite (or BRST) cohomology of the
$\cW_3$ algebra -- carries the structure of a G-algebra. This G-algebra
has a quotient which is isomorphic to the G-algebra of polyvector fields
on the base affine space of $SL(3,\CC)$.  Details will appear elsewhere.}{}

}
\textheight=7.8truein
\setcounter{footnote}{0}
\renewcommand{\thefootnote}{\alph{footnote}}

%
%
\section{Introduction and preliminaries}
\noindent

  W-gravity models are reparametrization-invariant field theories in which
the vector-generated diffeomorphism symmetry is extended by higher tensor
structures.  For example, while a scalar field transforms under an
infinitesimal diffeomorphism by the vector $\xi^\mu(x)$,
$x'^\mu = x^\mu + \xi^\mu(x)$, as
$\delta \phi(x) = \xi^\mu(x) \partial_\mu \phi(x)$, in a W-gravity
there would be additional invariances which at the linearized level act as
$$
\delta \phi(x) = \xi^{\mu_1 \cdots \mu_n}(x)
 \partial_{\mu_1} \cdots \partial_{\mu_n} \cdots \phi(x)
 $$
for some integer $n>1$.  Just as diffeomorphisms are associated
with massless particles of spin two, {\it i.e.} metric fluctuations,
so also W-gravities lead to theories of massless higher-spin
fields.   The goal of constructing a non-trivial
interacting QFT with massless higher-spin fields in four dimensional
spacetime has proved to be rather elusive, so it is not surprising that
very little is known about W-gravities there.

  Quite recently, progress has been made on the problem of constructing
$\cW$-gravities in two dimensions.    Classically such models are somewhat
trivial, since -- as for ordinary 2-d gravity -- a simple counting shows that
there are no propagating gravitational degrees of freedom after accounting for
gauge invariances (see \cite{Hull} and references therein).
However, upon quantization there are two possibilities
which are both of interest: either the decoupling of the $\cW$-gravitational
degrees of freedom is maintained, providing a generalization of the
world-sheet reparametrization invariance in the first-quantized description of
string theory and thus possibly new critical string theories; or the
decoupling is not maintained, and some $\cW$-gravitational fields become
propagating due to quantum corrections as in the so-called off-critical
theories.  In either case, these models provide a mathematical physics
laboratory to study the fascinating problem of quantizing
a gauge theory based on a non-linear algebra of constraints -- a
$\cW$-algebra extension of the Virasoro algebra.

  It is this algebraic structure which provides a definition of $\cW$-gravity
models even though the associated $\cW$-geometry is not yet well understood.
In  particular, by analogy with the standard DDK ansatz for 2-d
gravity\cite{David} \cite{DK}, there exists a well-motivated BRST-quantization
of $W_3$-gravity coupled to conformal matter with central charge less than or
equal to two\cite{TM,BLNW,BMPa}.  So, there is a nilpotent BRST charge $d$,
$d^2 = 0$, and the physical state subspace is defined by the condition
$d |phys\rangle = 0$, modulo the equivalence relation
$|phys\rangle \sim |phys\rangle + d |\chi\rangle$.  In other words,
a physical state is a representative of a BRST cohomology class.
Although (unlike the DDK case) calculating the physical spectrum
of these $\cW$-gravity models seems extremely difficult,
it is reasonable to expect that the study of this cohomology space
will lead to a better understanding of (quantum) $\cW$-geometry.

 In this paper I partially summarize work on this problem done in
collaboration with Peter Bouwknegt and Krzysztof Pilch.  I restrict the
discussion to one model: $\cW_3$-gravity coupled to a matter sector
consisting of two scalar fields ($c^M=2$).
In the corresponding string theory these scalar fields would embed the world
sheet of the string into a two dimensional space-time.  But, moreover,
since this is a non-critical theory there are dynamical gravitational degrees
of freedom which under the DDK-type ansatz are described by a pair of
scalar fields of ``wrong sign'' with a background charge source, the so-called
Liouville sector.  Thus, in this string language,
the model describes a $2+2$ dimensional string in non-trivial background
fields.  In a conformal gauge quantization of this string
the calculation splits into almost decoupled left- and right-moving sectors,
the physical states are then computed from the BRST cohomology of a
tensor product of two scalar field Fock space modules of the $\cW_3$ algebra.

  In Section 2 I present more clearly the cohomology problem which
is required to compute the physical spectrum of
the $D = 2+2$ $\cW_3$ string.  The complete result for this
cohomology is given at the start of Section 3, followed by some
justification for a special set which is picked out in
analogy with the ``Seiberg bound'' of the DDK case \cite{Seib}.
I then discuss in Section 4 the structure of the corresponding
operator cohomology as a Gerstenhaber (G-) algebra.   In particular,
the special case selected above is shown to be geometrically
modeled by the G-algebra of regular polyvector fields on the base affine
space of $SL(3,\CC)$.   For more details the reader is
referred to the literature\cite{BMPa} \cite{BMPb} \cite{BMPc}
and the forthcoming paper\cite{BMPd}.

 Throughout this paper I will use the notation $\bfh$ for
the Cartan subalgebra, $\bfh^*_\ZZ$ for the set of integral weights,
$P_+$ for the set of dominant integral weights,
$P_{++}$ for the set of strictly dominant integral weights.

\newpage

%
\section{Fock space BRST complex for the $\cW_3$ algebra}

The $\cW_3$ algebra with central charge $c\in\CC$
 (see {\it e.g.\ } the
review on $\cW$-algebras\cite{BS}, and references therein)
is generated by the
operators $L_n$ and $W_n$, $ n\in\ZZ$,
\begin{eqnarray}
\left[ L_m,L_n\right] &=& (m-n)L_{m+n}+
\textstyle{c\over 12}m(m^2-1)\de_{m+n,0} \nonumber \\
\left[ L_m,W_n\right] &=& (2m-n) W_{m+n}   \nonumber \\
\left[ W_m,W_n\right] &=&   (m-n)\left( \textstyle{1\over 15}
(m+n+3)(m+n+2)-\textstyle{1\over 6}
(m+2)(n+2)\right)L_{m+n} \nonumber \\
&& {}+  \be (m-n) \La_{m+n} +\textstyle{c\over 360}m(m^2-1)(m^2-4)\de_{m+n,0}
	     \,,
\label{eqAa}
\end{eqnarray}
where $\be=16/(22+5c)$ and
$$
\La_{m}=\sum_{n\leq-2} L_{n} L_{m-n} +  \sum_{n>-2} L_{n} L_{m-n}
  -\textstyle{3\over 10}(m+3)(m+2) L_m\,.
\label{eqAb}
$$
As for Lie algebras the commutators of generators satisfy the Jacobi
identities, but the $\cW$-algebras are clearly distinguished from
Lie algebras by their nonlinear structure.
The generators of the algebra $\cW_3$ may be decomposed according to
the $(-{\rm ad}\,L_0)$ eigenvalue,
$\cW_+ \oplus \cW_0 \oplus \cW_-$, where
$\cW_{\pm}=\{L_n,W_n\,|\, \pm n>0\}$
and the Cartan subalgebra $\cW_0$ is spanned by $L_0$ and $W_0$.
This is not a usual triangular decomposition with respect to $\cW_0$ --
in particular, $({\rm ad}\,W_0)$ is not diagonalizable.

  Despite these differences, a non-trivial BRST cohomology may
be defined for $\cW_3$ by analogy with that for the Virasoro algebra.
Corresponding to the two sets of generators $L_m$ and $W_m$, $m\in \ZZ$,
introduce two sets of ghost oscillators $(b^{[i]}_m,c^{[i]}_m)\,,i=2,3$ with
anticommutation relations
$\{c^{[i]}_m ,b^{[i']}_n\}= \delta^{i i'} \delta_{m+n,0}$.
The ghost Fock space $F^{\rm gh}$ is graded by the ghost number
${\rm gh}(\cdot)$, where ${\rm gh}(c^{[i]}_n)=-{\rm gh}(b^{[i]}_n)=1$,
while the ghost
number of the ($sl(2,\CC)$-invariant) vacuum state is zero.
For any two positive energy\footnote{A positive energy
module has $L_0$ diagonalizable
with finite dimensional eigenspaces, and with the
spectrum bounded from below.}  $\, \cW$-modules $V^M$ and $V^L$, such that
$c^M + c^L = 100$, there exists a complex $(V^M \otimes V^L \otimes
F^{{\rm gh},n}, d)$, graded by ghost number, and with a differential
(BRST operator) $d$ of degree $1$,
\begin{eqnarray}
d
&=& \sum_{m}\big( c^{[3]}_{-m}(\widetilde W^M_m - i \widetilde W^L_m) +
		c^{[2]}_{-m}(L^M_m+L^L_m) \big) \nonumber \\
&& {} + \sum_{m,n}\big(
-\half(m-n)\,c^{[2]}_{-m}c^{[2]}_{-n}b^{[2]}_{m+n}
-(2m-n)c^{[2]}_{-m}c^{[3]}_{-n}b^{[3]}_{m+n} \nonumber \\
&& \quad\quad\quad\,\,
-\textstyle{1\over 6}\mu(m-n)(2m^2-mn+2n^2-8)\,
c^{[3]}_{-m}c^{[3]}_{-n}b^{[2]}_{m+n}\big) \nonumber \\
&& {} + \sum_{m,n,p}\big(-\half(m-n)\, c^{[3]}_{-m}c^{[3]}_{-n}b^{[2]}_{-p}
(L^M_{m+n+p}-L^L_{m+n+p})\big)\,,
\label{bbrst}
\end{eqnarray}
where $\widetilde W^{M,L}=W^{M,L}/\sqrt{\be^{M,L}}$ and
$\mu=(1-17\be^M)/(10\be^M)$.
The cohomology of $d$ at  degree $n$ will be denoted by
$H^{n}(\cW,V^M\otimes V^L)$, and called the BRST cohomology of the $\cW_3$
algebra on $V^M\otimes V^L$.

 In this paper I will discuss the
case in which both $V^M$ and $V^L$ are free scalar field Fock space
modules of the $\cW$-algebras.  Let $\ph^i(z)$, $i=1,2$ be two free scalar
fields normalized such that
$\ph^i(z)\ph^j(w) = -\de^{ij}\ln(z-w) + \, {\rm regular}$,
and coupled to a background charge $\al_0\rh$, where $\rh = \al_1 + \al_2$ is
the principal vector of the Lie algebra $s\ell_3$.  In other words, the
stress-energy tensor, $T(z)$, has the ``improved'' form
$$
T(z)  = -\half (\p\ph(z),\p\ph(z)) - i\al_0 (\rh,\p^2\ph(z))\,.
\label{eqBFT}
$$
One easily checks from Wick's theorem that
$T(z) T(w) = {c(\al_0)/2\over {(z-w)^4}} + {2\over {(z-w)^2}} T(w) +
{1\over {z-w}} \p T(w) + \, {\rm regular}$, where $c(\al_0)=2 -24\al_0^2$ .
Decomposing in modes, $T(z) = \sum L_n z^{-n-2}$, this is equivalent to
the statement that the $L_n$ generate a Virasoro algebra of central charge
$c(\al_0)$.  The extension to a realization of the full
$\cW_3$ algebra is similarly encoded in the spin-3 chiral current ( after
a choice of orthonormal basis with respect to which
the simple roots of $s\ell_3$ are $\al_1 = (\sqrt{2},0)\, , \,\,
\al_2 = (-{1\over{\sqrt{2}}},{\sqrt{3}\over{\sqrt{2}}})$)
$$
\begin{array}{lcl}
{1\over \sqrt{3 \be}} W(z)  &=& {-i\over {3 \sqrt{6}}}\,
(\p\ph^1\p\ph^1\p\ph^2 - \p\ph^2\p\ph^2\p\ph^2) +  \\
& & {} \al_0 \, ({1\over 2} \p\phi^1 \p^2\phi^1 +
{1\over \sqrt{3}} \p\phi^2 \p^2\phi^1 -
{1\over 2} \p\phi^2 \p^2\phi^2) +
{i\over {2 \sqrt{2}}} \al_0^2\,
(\p^3\phi^1 - {1\over \sqrt{3}} \p^3\phi^2) \, .
\end{array}
\label{eqBFW}
$$

  The expansion of $\phi^j$ into modes is given by
$$
\phi^j(z) = \phi_0^j - i\,p^j\,{\rm log}z  +
i \sum_{n \neq 0}\al_n^j\, z^{-n} \, .
$$
Let $\cF(\La,\al_0)$ denote the Fock space of these $2$ scalar fields, where
$\La$ -- the ``momentum'' vector -- labels the Fock space vacuum
$|\La\rangle$ defined by $p^i|\La\rangle =\La^i |\La\rangle$ and
$\al_n |\La\rangle = 0\, , \,\, n>0$.  As a $\cW_3$ module the
Fock space is highest-weight, with
\begin{eqnarray}
L_0 |\La\rangle &=& h(\La)|\La\rangle = \half (\La,\La+2\al_0\rh)|\La\rangle
\nonumber \\
W_0 |\La\rangle &=& w(\La)|\La\rangle =
\sqrt{3\be}\,\th_1\th_2\th_3 |\La\rangle \, ,
\label{eqBFh}
\end{eqnarray}
where (the weights $\La_1$ and $\La_2$ are the fundamental weights of
$s\ell_3$)
$$
\th_1=(\La+\al_0\rh,\La_1)\,,\quad
	    \th_2=(\La+\al_0\rh,\La_2-\La_1)\,,\quad
	    \th_3=(\La+\al_0\rh,-\La_2)\,.
$$
It is important to note that both $h(\La)$ and $w(\La)$
in (\ref{eqBFh}) determine $\La$ only up to a
Weyl rotation $\La\rightarrow w(\La+\al_0\rh)-\al_0\rh$, $w\in W$.

  Finally, then, the case of interest here is
$\al_0^M=0$ and $-i\al_0^L =2$, {\it i.e.} $c^M=2, c^L=98$.
The Fock space momenta are {\it a priori} free, however it turns
out that without loss of generality the momenta may be restricted to
the $s\ell_3$ weight lattice.  Thus the cohomology to be discussed is
\begin{equation}
\cH \equiv \bigoplus_{ (\La^M,-i\La^L) \in \bfh^*_\ZZ \otimes \bfh^*_\ZZ}
H(\cW, F(\La^M,0) \otimes F(\La^L,2i) \, .
\label{eqBp}
\end{equation}

%
\section{Physical states of the $D=2+2$ $\cW_3$ string}

 The following result summarizes the solution\cite{BMPa} \cite{BMPb}
to the cohomology problem (\ref{eqBp}) posed in the previous section.
\begin{theorem} For the cohomology (\ref{eqBp}) with differential
(\ref{bbrst}),

(i) $\cH$ carries the structure of a
$\bfg\oplus\bfh$ module ($\bfg\cong s\ell_3$, $\bfh\cong u_1 \oplus u_1$).
This module is completely reducible under $\bfg\oplus\bfh$.

(ii) There exists a (non-canonical) isomorphism (as $\bfg\oplus
\bfh$ modules)
$$
\cH^i\ \cong\ \cH_{\rm pr}^i \oplus \cH_{\rm pr}^{i-1}
  \oplus \cH_{\rm pr}^{i-1} \oplus \cH_{\rm pr}^{i-2} \,.
$$
 The cohomology $\cH_{\rm pr}$ is isomorphic
(as a $\bfg\oplus \bfh$ module) to the direct sum of irreducible
modules $\cL(\La)\otimes \CC_{\La'}$ with momenta
$(\La,\La')\in \bfh^*_\ZZ \otimes \bfh^*_\ZZ$ lying in
a set of disjoint cones $\{ \cS^n_w + (\la,w^{-1}\la)\,|\,\la\in P_+\}$,
{\it i.e.}
$$
\cH_{\rm pr}^n\ \cong\ \bigoplus_{w\in W} \bigoplus_{(\La,\La')\in
  \cS_w^n} \bigoplus_{\la\in P_+} \left( \cL(\La+\la) \otimes
  \CC_{\La' + w^{-1}\la} \right) \,,
$$
where the sets $\cS_w^n$ (tips of the cones) are given in Table 1.
\end{theorem}

\medskip

\begin{tabular}{||c|c|c||} \hline
$n$ &  $w$  &  $\cS_w^n$ \\  \hline
$0$ & $1$ &  $(0,0)$  \\ \hline
$1$ & $1$ &  $(\La_1,-\La_1+\La_2)$, $(\La_1+\La_2,0)$,
             $(\La_2,\La_1-\La_2)$  \\
    & $r_1$ & $(0,-2\La_1+\La_2)$ \\
    & $r_2$ & $(0,\La_1-2\La_2)$ \\ \hline
$2$ & $1$ & $(2\La_1,-\La_1)$, $(0,-\La_1-\La_2)$,
            $(2\La_2,-\La_2)$ \\
    & $r_1$ & $(\La_1,-2\La_1)$, $(\La_2,-3\La_1+\La_2)$,
              $(0,-4\La_1+2\La_2)$ \\
    & $r_2$ & $(\La_2,-2\La_2)$, $(\La_1,\La_1-3\La_2)$,
              $(0,2\La_1-4\La_2)$ \\
    & $r_1r_2$ & $(0,-3\La_2)$ \\
    & $r_2r_1$ & $(0,-3\La_1)$ \\ \hline
$3$ & $1$ & $(\La_1+\La_2,-\La_1-\La_2)$ \\
    & $r_1$ &  $(\La_2,-2\La_1-\La_2)$, $(\La_1,-4\La_1+\La_2)$,
              $(\La_2,-5\La_1+2\La_2)$  \\
    & $r_2$ & $(\La_1,-\La_1-2\La_2)$, $(\La_2,\La_1-4\La_2)$,
              $(\La_1,2\La_1-5\La_2)$ \\
    & $r_1r_2$ & $(\La_2,-\La_1-3\La_2)$, $(0,\La_1-5\La_2)$,
                 $(\La_2,-5\La_2)$ \\
    & $r_2r_1$ & $(\La_1,-3\La_1-\La_2)$, $(0,-5\La_1+\La_2)$,
                 $(\La_1,-5\La_1)$ \\
    & $r_1r_2r_1$ & $(0,-2\La_1-2\La_2)$ \\ \hline
$4$ &  $r_1$ & $(0,-4\La_1-\La_2)$ \\
    & $r_2$ & $(0,-\La_1-4\La_2)$ \\
    & $r_1r_2$ & $(\La_2,-2\La_1-4\La_2)$, $(\La_1,-\La_1-5\La_2)$,
                 $(0,-6\La_2)$ \\
    & $r_2r_1$ & $(\La_1,-4\La_1-2\La_2)$, $(\La_2,-5\La_1-\La_2)$,
                 $(0,-6\La_1)$ \\
    & $r_1r_2r_1$ &  $(0,-3\La_1-3\La_2)$, $(2\La_1,-4\La_1-3\La_2)$,
                $(2\La_2,-3\La_1-4\La_2)$ \\ \hline
$5$ & $r_1r_2$ & $(0,-2\La_1-5\La_2)$ \\
    & $r_2r_1$ & $(0,-5\La_1-2\La_2)$ \\
    & $r_1r_2r_1$ &  $(\La_1,-5\La_1-3\La_2)$, $(\La_1+\La_2,-4\La_1-4\La_2)$,
        $(\La_2,-3\La_1-5\La_2)$ \\ \hline
$6$ & $r_1r_2r_1$ & $(0,-4\La_1-4\La_2)$ \\ \hline
\end{tabular}
\medskip

\centerline{\it Table~1.\ The sets $\cS_w^n$}\bigskip

The action of $\bfg$  is via the zero modes of the Frenkel-Kac-Segal vertex
operator construction of level 1 affine $su(3)$ (using only the
$\al_0^M =0$ matter fields of course),
while $\bfh$ acts as $-ip^L$ (with eigenvalues $-i\La^L$).
One easily checks directly that these are non-trivial operators which
commute with $d$.  That leaves the  computation in $(ii)$.  [Note that
the quartet structure in terms of so-called ``prime'' states is
a nontrivial result, and not an immediate consequence of a relation
between the full cohomology and some relative subcomplex.]

  For the Virasoro case\cite{BMPV} it was straightforward to work with the Fock
space oscillators.  The basic observation is well illustrated in calculating
the cohomology of $d_0 = c a^\dagger$ on the Fock space of a single oscillator,
$[a,a^\dagger] = 1$, $\{b,c\}=1$.  The only non-trivial cohomology state
is clearly the state $c |0>$ where $a|0> = b|0> = 0$
(any Fock space state of the form $c \psi\rangle$ is annihilated by $d$,
but on modding out the $d$-exact states only $c |0\rangle$ is non-trivial).
Thus one simply has to introduce a degree on the space of oscillators for
which the lowest degree term in $d$ has the form of $d_0$ for each oscillator.
The higher degree terms then act trivially and
the full cohomology is again just one dimensional.  A modification of this
idea completely solves the Virasoro case.  Unfortunately, for the
$\cW$ case it does not seem possible to assign a degree which is even close to
working in the same way, and one must find other techniques (there
are alternative discussions for the Virasoro case\cite{LZa}, see
also\cite{BMPgc}).

 As mentioned in the introduction, for this paper I will only
discuss the justification of $(ii)$ in the case that
$-i\La^L + 2 \rho \in P_+$, which is essentially $w=1$ in
the above.  The other sectors are deduced from a conjecture based
on generic results for $c^M < 2$, together with the $w=1$ sector result,
and there is not enough space here to motivate this conjecture properly.

\subsection{The case $-i\La^L + 2\rho \in P_+$}

  The $d_0 = c a^\dagger$ cohomology for a single oscillator,
as discussed above, was simple to calculate because the Fock module has
a particularly simple structure -- it is free over $a^\dagger$ -- which
``fits'' the form of $d_0$.  For the  $\cW$-algebra computation $d$ has the
form ``$d \sim c^{[2]}_{-n} L_n + c^{[3]}_{-n} W_n + \cdots$'', so it is
natural to calculate the cohomology where $V^M$ an $V^L$ are modules free over
part of the $\cW_3$-algebra.  A Verma module $M(h,w,c)$ is defined as the
module induced by $\cW_-$ from an eigenstate of $\cW_0$, $|h,w\rangle$,
({\it i.e.} free over $\cW_-$ with a single generator).  The contragradient
Verma module is denoted by $\ffM(h,w,c)$.   A slight generalization
arises naturally in considering the submodule structure of a Verma module:
Due to the non-diagonalizability of $W_0$ it is useful to define
the {\it generalized Verma module}, $M(h,w,c)_N$, as the
positive energy module induced from an $N$-dimensional
indecomposable representation of $\cW_0$.
Let $v_0,\ldots,v_{N-1}$ be a canonical basis such that
\begin{eqnarray}
L_0 v_i &=& hv_i\,,\quad  i=0,\ldots,N-1\,, \nonumber \\
W_0v_0 &=& wv_0\,,\quad W_0 v_i =wv_i+v_{i-1}\,,\quad
i=1,\ldots,N-1\,.
\label{eqlowo}
\end{eqnarray}
Then $M(h,w,c)_N$ is spanned by the monomials of the form
$$
L_{-n_1}\ldots L_{-n_i}W_{-m_1}\ldots W_{-m_j} v_k\,,\quad
i,j\geq 0 \,,\quad k=0,\ldots,N-1\,,
$$
on which the generators in $\cW_-$ act freely, while the action of those in
$\cW_+$ and $\cW_0$ is  determined using (\ref{eqAa}), (\ref{eqlowo}), and
$$
L_nv_i=W_nv_i=0\,,\quad n>0\,,\quad i=0,\ldots,N-1\,.
$$
For $N=1$ the usual definition of the Verma module
$M(h,w,c)$ is clearly recovered.

Taking $V^M$ and $V^L$ as contragrediently-related (generalized) Verma
modules, the cohomology of $d$ is quite easily calculated via the introduction
of an appropriate degree.
\begin{lemma} $\, $ \\
The cohomology
$H(\cW,M(h(\La^M),w(\La^M),c(\al_0^M))_N \otimes
\ffM(h(\La^L),w(\La^L),c(\al_0^L)))$ is non-vanishing if and only if
$-i(\La^L + \al_0^L \rho) = w(\La^M + \al_0^M \rho)$ for some
$w \in W$, in which case it is spanned by the states
$v_0$, $c_0^{[2]} v_0$, $c_0^{[3]} v_{N-1}$ and
$c_0^{[2]} c_0^{[3]} v_{N-1}$, where
$v_i = v_i^M \otimes \bar{v}^L \otimes |0\rangle_{gh}$.
\end{lemma}

  This result still seems quite far from the Fock space calculation required,
however there are three observations on Fock space modules $F(\La,\al_0)$ which
make it immediately applicable.  The first two are reasonably standard to prove
using the free field realization and the existence of a Hermitian inner product
for $c=2$.\
\begin{lemma} $\, $ \\
(i) For $-i\La^L + 2 \rh \in P_+$ there is
an isomorphism $F(\La^L,2i) \cong \ffM(h(\La^L),w(\La^L),98)$.

\noindent
(ii) For $c=2$,
the Fock space $F(\la,0)$ is completely reducible. Explicitly,
for all $\la\in \bfh^*_\ZZ$, the Fock spaces decomposes as
$$
F(\la,0)\ \cong\ \bigoplus_{\La\in P_+}\ m^\La_\la\, L(h(\La),w(\La),2)\,,
$$
where $m^\La_\la$ is equal to the multiplicity
of the weight $\la$ in the irreducible
finite dimensional representation $\cL(\La)$
of $s\ell_3$ with highest weight $\La$.
\end{lemma}
\smallskip

\noindent
The last observation\cite{BMPb} is that for $c=2$ one may find  resolutions
of $L(\La)$ in terms of (generalized) Verma modules $M(h,w,c=2)_N$,
{\it i.e.} there is a complex (of finite length),
$$
0 \rightarrow \cC^0  \rightarrow  \cC^1 \rightarrow \cdots   \, ,
$$
where the $\cC^i$ are a finite direct sum of generalized Verma modules,
the maps are embeddings, and where the non-trivial cohomology
is at $\cC^0 = M(h,w,c=2)$ and is given by the irreducible representation
$L(h,w,c=2)$ (a slight misprint\cite{BMPb} in the resolution of $L(\La)$ for
$\La \in P_{++}$ has been corrected\cite{BMPd}).
This result is, strictly speaking, conjectural, but has been
tested by quite exhaustive Mathematica$^{\rm TM}$ calculation.

  Figuratively speaking, the following steps have been applied to this stage
(using $H_d(\cdot)$ to denote $H(\cW,\cdot)$, and $H_\delta$ to denote the
cohomology in the resolution of the irreducible representation as discussed
above):
$$
H_d(F\otimes F) = \oplus H_d(L\otimes \ffM) =
\oplus H_d H_\delta (M_N \otimes \ffM) =
\oplus H_\delta H_d(M_N \otimes \ffM) \, .
$$
The cohomology on the left is the required Fock space cohomology,
while that on the right is known from
Lemma 1.  The exchange of orders in calculating the cohomology is allowed,
basically because after grading by $L_0$ every computation
can be reduced to one on finite dimensional vector spaces.
Thus, the Fock space cohomology for $-i\La^L + 2 \rho \in P_+$
may now be rather simply re-assembled by simply going backwards in these
steps.

\section{The operator algebra}

  In this final section I will try to give the flavour of the type of result
which can be obtained on the operator algebra.  Analogous results
for the Virasoro case may be found in \cite{eb} \cite{WuZu} \cite{LZ}.
I will not attempt to give a complete result since that requires quite a
deal of mathematical preliminaries.

 Given a state $|\cO\rangle$ in the BRST complex, one may assign an operator
$\cO(z)$ such that
$$
|\cO\rangle = \lim_{z \rightarrow 0} \cO(z) |0\rangle \, .
$$
The operator product expansions of these operators are graded commutative
(graded by ghost number) and associative -- these are just statements
about free fields.  Perhaps the only subtlety is the existence of
``phase-cocycles'', which are required to ensure mutual locality.

  To explain this, let me consider as an example the
operator corresponding to the Fock space state
$|\La^M,\La^L\rangle$, which is naively the normal-ordered exponential
$$
V_A(z) = :e^{i (\La_A^M, \phi^M(z)) + i (\La_A^L,\phi^L(z))}: \, .
$$
It is easily verified that
\begin{equation}
V_A(z) V_B(w)  = (z-w)^{h_{AB}}
           :e^{i (\La_A^M, \phi^M(z)) + i (\La_A^L,\phi^L(z)) +
               i (\La_B^M, \phi^M(w)) + i (\La_B^L,\phi^L(w))}: \, ,
\label{vv}
\end{equation}
where $h_{AB} = (\La_A^M, \La_B^M) + (\La_A^L, \La_B^L)$,
as an operator product expansion -- {\it i.e.} for $|z|>|w|$ the product
is just composition of operators, but for the remainder it is defined by
analytic continuation.  However, the phase introduced by the $(z-w)$ factor
will not in general be consistent with (\ref{vv}), analytic continuation, and
$V_A(z) V_B(w) = V_B(w) V_A(z)$.  To account for the phase, one replaces
the exponential by
$$
\cV_A(z) = V_A(z) e^{i \pi \xi_A^M \cdot p^M + i \pi \xi_A^L \cdot p^L} \, ,
$$
where the phase-cocycles are chosen so that
$$
 \xi_A^M \cdot \La_B^M + \xi_A^L \cdot \La_B^L -
 \xi_B^M \cdot \La_A^M - \xi_B^L \cdot \La_A^L =
 \La_A^M \cdot \La_B^M + \La_A^L \cdot \La_B^L \quad ({\rm mod}\,\, 2) \, .
$$
Thus, the problem of constructing an operator product algebra basically reduces
to finding such phase-cocycles -- which is relatively straightforward.
In the present case, this requires restriction of the momenta
$(\La^M,-i\La^L)$ to the lattice $L$, where
$$
L\ \equiv\ \{ (\la,\mu) \in \bfh^*_{\ZZ} \otimes \bfh^*_{\ZZ}\,|\,
  \la - \mu \in \ZZ\cdot\De_+ \} \,.
$$
This restriction is already required to produce a
vertex operator algebra whose operator
product expansion with the cohomology is meromorphic.
I will denote the operator product algebra corresponding to the BRST
complexes based on Fock spaces with weights in the lattice $L$ by $\cA_L$.
Indeed to construct the analogue of the cohomology problem (\ref{eqBp})
the direct sum should, strictly speaking, be restricted to weights
in the lattice $L$.  As is clear from the results in Section 3
this does not involve any loss of generality.

  The BRST operator $d$ acts on $\cA_L$ by commutator -- equivalently,
let $J(z)$ be the BRST current, then the action of $d$ is given by
$$
\oint_{C_z} {dx\over {2\pi i}} J(x) \, \cO(z) \, ,
$$
where the contour $C_z$ surrounds the point $x=z$ counterclockwise.
The operator cohomology, $\cH_{op}$, is isomorphic to the physical states
of Section 2.  When I talk about an operator in $\cH_{op}$ I will mean an
element of $\cA_L$ which is a representative of operator cohomology in
the same sense that a physical state is a representative of
$H(\cW,F^M\otimes F^L)$.
There is a Virasoro algebra acting on $\cA_L$ which descends to $\cH_{op}$,
the generators given by $L_n = \{d,b^{[2]}_n\}$.  Since $L_0$ is diagonal on
$\cA_L$, clearly only the subspace annihilated by $L_0$ can be non-trivial
in $\cH_{op}$.  The algebra $\bfg\oplus\bfh$ of Theorem 1
also acts on $\cA_L$ and descends to $\cH_{op}$.

  The aim now is to give a clean description of the algebra of
operators in $\cH_{op}$.  It is a rather general result for gravity models
\cite{LZ} that on $\cH_{op}$ there is a graded-commutative associative
product, $\cdot$, and a bracket, $[\, ,\, ]$, with respect to which $\cH_{op}$
is a Lie superalgebra.  The dot product of two operators in $\cH_{op}$ is
given by (note that in this notation the bracket does {\it not} denote the
commutator)
$$
(\cO \cdot \cO')(z) = {1\over {2\pi i}}\oint_{C_z}{dx\over {x-z}}
\cO(x)\, \cO'(z) \, .
$$
Since all non-trivial cohomology states are annihilated by $L_0$, any singular
terms in the operator product expansion on the rhs are trivial in operator
cohomology.  The associativity and graded commutativity of the product at the
level of cohomology follow immediately.  The bracket is given by
$$
[\cO,\cO'](z) = \oint_{C_z}{dx\over {2 \pi i}}\,
(b^{[2]}_{-1}\cO)(x) \, \cO'(z) \, ,
$$
which is BRST invariance since $d b^{[2]}_{-1} \cO = L_{-1}\cO$, and
the action of $L_{-1}$ on a field in $\cA_L$ is equivalent to taking
its derivative.   The grading of the bracket, as well as the graded-Jacobi
identity can be checked, along with the statement that the bracket
acts as a superderivation of the dot-algebra.
Together with grading, and the structures of dot and bracket, these properties
identify $\cH_{op}$ as a Gerstenhaber (or G-) algebra\cite{Ge}.

  The archetypal example of a G-algebra is the algebra of polyvector
fields on a smooth manifold $M$, $\cP(M)$, graded by the degree of the
polyvector.  More precisely, at the zeroth level in grading is just the
commutative algebra (under point-wise multiplication) of polynomial functions
on $M$, $\cA$.  At degree one are polynomial-valued vector fields on $M$,
which of course act as derivations on the functions.  The bracket at this
level can be taken as the Lie bracket of vector fields, extended to functions
in the usual way.  At degree two are antisymmetric derivation-valued
derivations of $\cA$ -- {\it i.e.} bi-vectors on $M$.  This construction
clearly continues up to degree equal to the dimension of $M$.
The dot product is just the wedge product on polyvectors,
and the bracket is defined inductively from the Lie bracket.  An obvious
aim would be to understand the G-algebra of operator cohomology in this
geometrical setting.  The following gives exactly such a result.

\begin{theorem}\cite{BMPd}

\noindent
(i)  The ghost number zero subspace $\cH_{op}^0$ is a subalgebra, and
as an $s\ell_3$ module it decomposes as
$\cH_{op}^0 \cong \bigoplus_{\La\in P_+} \cL(\La)$.  Thus $\cH_{op}^0$ is
a  ``model space'' for $s\ell_3$, and in fact is isomorphic
as an algebra to the usual geometrical model -- namely $\cP^0(A)$,
the space of polynomial functions on the base affine space
$A\equiv N_+\backslash SL(3)$ \cite{BGG}.

\noindent
(ii)  This isomorphism extends to a G-algebra homomorphism
$\pi : H \rightarrow \cP(A)$, and there exists an ideal
$\cI \subset \cH$ such that
$$
\matrix{
0 & \mapright{} & \cI & \mapright{} & \cH & \mapright{\pi} &
\cP(A) & \mapright{} & 0\\}
$$
is an exact sequence of G-algebras.
\end{theorem}

\noindent
The proof of $(i)$ is largely direct calculation.  The $s\ell_3$
decomposition of $\cH_{op}^0$ may be read from Theorem 1.
Notice that the base affine space for $s\ell_3$ is isomorphic to \cite{Zh}
$\CC\,[x^i,y_i] / \langle x^iy_i \rangle$,
{\it i.e.} the space of polynomials in the $6$ variables $x^i, y_i$
modded out by a single relation $x^iy_i=0$, where $x^i$ and $y_i$
transform in the $\bf 3$ and $\bar{\bf 3}$ of $s\ell_3$ respectively.
Denote the representatives of $\cH_{op}^0$ which transform in these
representations by $\Psi_{x^i}$ and $\Psi_{y_i}$.  As discussed
earlier, in cohomology these operators generate under dot product a
commutative and associative algebra.  Denote this algebra by $\cR$.
One easily sees from Theorem 1 that addition of Liouville momenta is
inconsistent with a singlet in the product of these representatives --
alternatively one can directly calculate $\Psi_{x^i}\Psi_{y_i}$ and show
that it is $d$-trivial.   So, a given monomial in the algebra generated
by these operators,
$$
\Psi_{x^i_1} \cdot\dots\cdot \Psi_{x^i_m}\cdot
\Psi_{y_j^1} \cdot\dots\cdot \Psi_{y_j^n}
$$
transforms under $\bfg \oplus\bfh$ exactly as the tensor product
representation, and thus decomposes either by contraction -- which by
associativity and the calculation mentioned above must vanish in
cohomology -- or by antisymmetrization which by commutativity clearly
vanishes.  Thus the only $s\ell_3$ representation which can possibly survive
is precisely the ``top" one (consistent with Theorem 1), and if
all dot products of the highest weight operators $\Psi_{x^1}$ and
$\Psi_{y_3}$ are non-trivial in cohomology then the algebra
${\cal R}$ is isomorphic to $P^0(A)$ as wanted.
Since these two operators are $s\ell_3$ highest weight,
they have Liouville momenta determined by $-i\La^L = \La^M$.  Hence they
have the form
\begin{eqnarray}
\Psi_{x^1} &=& P_{x^1} e^{i \La_1 \cdot \phi^+} \nonumber \\
\Psi_{y_3} &=& P_{y_3} e^{i \La_2 \cdot \phi^+} \, , \nonumber
\end{eqnarray}
where $\phi^\pm = \phi^M \pm i \phi^L$, and where the prefactor
$P$ denotes a level 2 operator built from the ghosts, $\partial \phi^\pm$,
and derivatives thereof. The precise expressions for the prefactors are not
needed, just the fact that for any representative in cohomology these
prefactors must contain a term quadratic in $i\partial \phi^-$ -- and indeed
an independent combination in each prefactor.  Such terms,
monomials in the prefactor built purely from $i\partial \phi^-$, will be
referred to as ``leading order'', and for these highest weight operators
one finds
\begin{eqnarray}
P_{x^1}|_{\rm leading} &=& {3\over 2} \big( (i\p\phi^{-,1}) (i\p\phi^{-,1})  +
   \sqrt{3} i\p\phi^{-,1} i\p\phi^{-,2} \big) \nonumber \\
P_{y_3}|_{\rm leading} &=& {3\over 4} \big( (i\p\phi^{-,1}) (i\p\phi^{-,1})  -
         3 \big( (i\p\phi^{-,2}) (i\p\phi^{-,2}) \big) \nonumber \, .
\end{eqnarray}
Consider now the dot product of $m$ $\Psi_{x^1}$'s and $n$ $\Psi_{y_3}$'s,
$$
\Psi_{x^1}\cdot \dots \Psi_{y_3}
= P_{mn} e^{i (m \La_1 + n \La_2) \cdot \phi^+} \, ,
$$
where the prefactor $P_{mn}$ is a level $2(m+n)$, ghost number zero operator
built as before.  There are no contractions between the purely-$\phi^+$
exponentials in computing the dot product.  There will be contractions
between the prefactors and the exponentials, and between prefactors.
A moment's thought shows that these cannot give rise to leading terms:
only the first non-singular term (from the Taylor series expansion
of the residue) contributes to the dot product, and this necessarily either
has higher derivatives, or non-zero powers of $i\partial \phi^+$ from
differentiating the exponentials.   Thus there is a unique source, providing
a manifestly non-zero leading term, which is the contribution from
the no-contraction term.

  Hence, the dot product $\Psi_{x^1}^m \Psi_{y_3}^n$ has a non-zero leading
term, and a similar argument using the explicit form of $d$ shows that it
is non-trivial in cohomology.  This completes the proof of part $(i)$.

  The extension of this step to the complete G-algebra as in $(ii)$ requires a
more detailed understanding of polyvectors on base affine space,
and sufficient calculation to identify the generators in cohomology.
This may certainly be done, the reader may consult \cite{BMPd}.
Let me here just explain the homomorphism, $\pi$, following \cite{LZ}.
Since at the zeroth grade this is an isomorphism,
I will use $X$ to denote both a given monomial in $x^i$ and $y_i$ as well as
the corresponding cohomology representatives,
unless there is the possibility of confusion.   Now, since
$[\cH_{op}^1,X] \in \cH_{op}^0$, for all $X\in \cH_{op}^0$,
the G-algebra structure may be used to extend $\pi$ to
$\cH_{op}^1$ via
\begin{equation}
[ \pi(\Psi^{(1)}),X ] \equiv \pi([\Psi^{(1)},X]) \, .
\label{def}
\end{equation}
Notice that the square bracket denotes
the bracket structure in $\cH_{op}$ as well as in $\cP(A)$, there
will be no possibility of confusion due to the explicit
use of the projection, $\pi$.
Having determined this map for ghost number 1 it
may be extended it to ghost number 2, and so on, exactly as above.

  By construction the map $\pi: \cH_{op} \rightarrow \cP(A)$ is a dot-algebra
homomorphism, as is easy to show inductively.  Suppose it is
true up to some ghost number, then if $\Psi \cdot \Psi'$
is ghost number one higher (for any $\Psi,\Psi' \in \cH_{op}^*$),
\begin{eqnarray}
\left[ \pi(\Psi\cdot\Psi'),X\right] &=&
\pi(\left[\Psi\cdot\Psi',X\right]) \\
&=& \pi(\Psi\cdot \left[\Psi',X\right]) +
(-1)^{|\Psi'|} \pi(\left[\Psi,X\right] \cdot \Psi') \\
&=& \pi(\Psi)\cdot \pi(\left[\Psi',X\right]) +
   (-1)^{|\Psi'|} \pi(\left[\Psi,X\right]) \cdot \pi(\Psi') \\
&=& \left[ \pi(\Psi)\cdot \pi(\Psi'),X\right]
\end{eqnarray}
where the third equality uses the induction hypothesis.
To show that the map $\pi$ is actually a G-algebra homomorphism is
equally straightforward.  Note that for any $\Psi\in H^*$,
$$
\pi([\Psi,X]) \equiv [ \pi(\Psi),X ] \, ,
$$
simply by the definition (\ref{def}) .  Since $\pi$ is
a ring isomorphism this is also true if $X$ is replaced by
any operator in $H^0$.  Again suppose it is true up to some
ghost number, and that (for any $\Psi,\Psi' \in H^*$)
$[\Psi,\Psi']$ has ghost number one higher.  Then
\begin{eqnarray}
\left[ \pi(\left[\Psi,\Psi'\right]),X\right] &=&
\pi(\left[\left[\Psi,\Psi'\right],X\right]) \\
&=& \pi(\left[\Psi,\left[\Psi',X\right]\right]) -
(-1)^{|\Psi'|} \pi(\left[\left[\Psi,X\right],\Psi'\right])  \\
&=& \left[ \pi(\left[\Psi,X\right]), \pi(\Psi')\right] -
(-1)^{|\Psi'|} \left[ \pi(\Psi), \pi(\left[\Psi',X\right])\right] \\
&=& \left[ \left[ \pi(\Psi),X\right], \pi(\Psi')\right] -
(-1)^{|\Psi'|}\left[ \pi(\Psi), \left[\pi(\Psi'),X\right]\right] \\
&=& \left[\left[ \pi(\Psi), \pi(\Psi')\right],X\right] \, .
\end{eqnarray}
Thus this $\pi$ is in fact the homomorphism required. $\Box$

  As a final remark, it is worth noting that the general
construction of the G-algebra in string theories, as alluded to
above, leads to a very special subset; namely, the Batalin-Vilkovisky
(BV-) algebras\cite{Ko} \cite{Get} \cite{PS}, for which
there is a nilpotent second order derivation $\De$ such that
$$
[a,b] = (-1)^{|a|} \left( \De(a\cdot b) - (\De a)\cdot b -
  (-1)^{|a|} a\cdot(\De b) \right) \,,
$$
for any $a,b$ in the algebra.  The role of $\De$ is played by the
ghost zero mode $b_0^{[2]}$.  It is possible to show\cite{BMPd} that
$\cP(A)$ has a BV-algebra structure, and in fact the sequence in the
statement of Theorem 2 $(ii)$ is an exact sequence of BV-algebras
which splits as a sequence of $\imath(\cP(A))$ dot modules, where
$\imath:\cP(A)\rightarrow \cH$ is a dot algebra homomorphism such that
$\pi\circ\imath={\rm id}$.

%
%
\nonumsection{Acknowledgements}

In various calculations throughout this work the Mathematica package
{\tt OPEdefs} of C.~Thielemans\cite{Th} has been extremely useful.
I would like to thank the organizers of the conference
``Confronting the Infinite'' for the opportunity to present this work.
The conference celebrated the work of two outstanding mathematical
physicists, Bert Green and Angas Hurst, the title apparently {\it not}
referring to age but rather the capacity to suffer through seminars.
I hope they will find this little contribution amusing, but in any
case I look forward to having another go when we celebrate their next
milestone.

\vfil\eject

\def\AdM#1{Adv.\ Math.\ {\bf #1}}
\def\AnM#1{Ann.\ Math.\ {\bf #1}}
\def\AnP#1{Ann.\ Phys.\ {\bf #1}}
\def\CMP#1{Comm.\ Math.\ Phys.\ {\bf #1}}
\def\FAP#1{Funct.\ Anal.\ Appl.\ {\bf #1}}
\def\IJMP#1{Int.\ J.\ Mod.\ Phys.\ {\bf #1}}
\def\InM#1{Inv.\ Math.\ {\bf #1}}
\def\JGP#1{J.\ Geom.\ Phys.\ {\bf #1}}
\def\JPA#1{J.\ Phys.\ {\bf A{#1}}}
\def\JRAM#1{J.\ reine angew.\ Math.\ {\bf {#1}}}
\def\JSP#1{J.\ Stat.\ Phys. {\bf {#1}}}
\def\LEM#1{L'Enseignement Math\'ematique {\bf {#1}}}
\def\LMP#1{Lett.\ Math.\ Phys.\ {\bf #1}}
\def\LNM#1{Lect.\ Notes in Math.\ {\bf #1}}
\def\MPL#1{Mod.\ Phys.\ Lett.\ {\bf #1}}
\def\NPB#1{Nucl.\ Phys.\ {\bf B#1}}
\def\PLB#1{Phys.\ Lett.\ {\bf {#1}B}}
\def\PNAS#1{Proc.\ Natl.\ Acad.\ Sci. USA {\bf #1}}
\def\PRep#1{Phys.\ Rep.\ {\bf #1}}
\def\PRA#1{Phys.\ Rev.\ {\bf A{#1}}}
\def\PRB#1{Phys.\ Rev.\ {\bf B{#1}}}
\def\PRD#1{Phys.\ Rev.\ {\bf D#1}}
\def\PRL#1{Phys.\ Rev.\ Lett.\ {\bf #1}}
\def\PTP#1{Prog.\ Theor.\ Phys.\ Suppl.\ {\bf #1}}
\def\SMD#1{Sov.\ Math.\ Dokl.\ {\bf {#1}}}
\def\TAMS#1{Trans.\ Amer.\ Math.\ Soc.\ {\bf #1}}
\def\TMP#1{Theor.\ Math.\ Phys.\ {\bf {#1}}}
\def\UMN#1{Usp.\ Mat.\ Nauk\ {\bf #1}}
\def\hepth#1{({\tt hep-th/{#1}})}
\def\condmat#1{({\tt cond-mat/{#1}})}

\nonumsection{References}

\end{document}